# Comments on redefinition of SI units based on fundamental physical constants with fixed values

V.V. Khruschov

Centre for Gravitation and Fundamental Metrology, VNIIMS, Ozernaya st. 46, Moscow, 119361, Russia

**Abstract**

Advantages and disadvantages of fixation of fundamental physical constants' values for definitions of SI units are considered. The case with a new definition of the mass unit on the base of a fixed value of the Avogadro constant is studied in detail. Criteria on choosing of a optimum FPC set with fixed values for the redefinition of the SI units are suggested. The minimal optimum FPC set that is consistent with the criteria is presented. The set comprises the speed of light, the constant of the ground state hyperfine transition of the caesium-133 atom, the Avogadro constant, the mass of the carbon-12 atom and the absolute magnitude of the electron charge. Comment on the redefinition of the kelvin is also made.

___________________________________________________________________________

1. **Introduction**

The problem of new definitions of a few basic SI units is under consideration now. It is proposed that new definitions will be based on fundamental physical constants (FPC) [1-5]. In consequence of this the practical quantum standards based on quantum Josephson and Hall effects might be incorporated in the SI, while the International Prototype of the Kilogram (IPK) will be eliminated out of the SI measure standards.

It is known the IPK possesses of imperfections, which have influence on the definitions of three SI units – the ampere, the mol and the candela – because of these definitions are used of the IPK. The main IPK imperfection consists in the uncontrollable change of its scale due to contamination and deterioration as it was obtained after comparing to its copies. In 1990 this change was about $5\times10^{-8}$ kg for 100 years. The IPK can be damaged; it is available only in one place, so it is difficult to pass its scale to other standards. The large density of the platinum-iridium alloy of the IPK and six its copies leads to considerable loss of precision when they are compared in air with steel standards.

Now the new definitions of some SI units were proposed, taking into account the imperfections, which were connected with using of the artifacts. The new definitions are based on FPC [1, 3, 6, 7, 8, 9, 10] and will aid in increasing of precision, stability and reproducibility of results of measurements at all times and in any place. The reason is that the units, which based on FPC, become independent from many effects of natural environment and human activity. An example is the redefinition of the measurement standard of frequency/time, that alloys for high precision of reproducibility of the frequency/time unit. In 1967 XIII General Conference on Weights and Measures (CGPM) defines the cesium second. In 1983 XVII CGPM defines so-called the "light metre". The definition of the SI length unit with the help of the light metre is the example of FPC fixation for metrological objective.

Notwithstanding the fact that the basic SI units do not coincide with the FPC, they can be represented through derived quantities of FPC [11]. As this takes place, any method of realization of the SI units in fact is connected with some FPC set, that is the basic set for the method.

The existing general structure of SI is consistent today and, possible, will be consistent with enquiries of metrological and all scientific community in the future. So the problem is finding an optimum set of fixed FPC for redefinition of the basic SI units.

In the present paper we propose the minimal FPC set, which has not taken into consideration so far. Using suggested definitions of SI units as examples [1, 2, 3, 8, 12] some advantages and disadvantages of FPC fixation are considered. The new FPC set is obtained in view of recent experimental advances in precise FPC measurements, along with supplementary restrictions for the FPC set. It is conceivable that these restrictions can be suitable for use in deciding on new definitions of SI units.

## 2. Redefinitions of SI units on the basis of fixed FPC values

It is well known that in 2005 the proposition was made of a redefinition of the SI mass unit based on the fixed Planck constant or the Avogadro constant as soon as possible [1]. In that year the International Committee for Weights and Measure (CIPM) adopted the recommendation on the preparative steps towards new definitions of the kilogram, the ampere, the kelvin and the mol in terms of fundamental constants [2].

Several versions of new definitions of mentioned above four basic units had been proposed, which were connected with four fixed FPC: the Planck constant $h$ (the kilogram), the elementary electric charge $e$ (the ampere), the Boltzmann constant $k_B$ (the kelvin) and the Avogadro constant $N_A$ (the mol), in addition to the fixed previously speed of light $c$ (the metre) [3]. Clearly, a putting into use definition of a unit must bring to no discontinuity of existing and new values for this unit. So the fixed values of $h$, $e$, $k_B$, and $N_A$ are bound to be into the definite limits prescribed by recommendations of metrological organizations.

There exist two new definitions of the mass unit, which are based on the use of the Planck constant and the Avogadro constant. It is desirable that the relative standard uncertainty of the transmission of the new mass unit is at $10^{-9}$ and the instability of the unit is less than $5 \times 10^{-10}$ kg per year for the compatibility with the IPK and the succession with the obtained previously results of mass measurements [12]. However, at present the values of the Planck constant and the Avogadro constant have been measured with inadequate precision.

According to the adopted recommendations and resolutions of the CIPM and the XXIII CGPM it is essential that the relative standard uncertainties for the $h$ and $N_A$ values be less than $2 \times 10^{-8}$ [2, 5]. This condition can be reduced, as the Consultative Committee for Mass and Related Quantities (CCM) recommends in 2010 [13], as follows:

– at least three independent experiments, including work both from watt balance and from International Avogadro Coordination projects (IAC), yield values of the relevant constants with relative standard uncertainties not larger than 5 parts in $10^8$. At least one of these results should have a relative standard uncertainty not larger than 2 parts in $10^8$.

– for each of the relevant constants, values provided by the different experiments be consistent at the 95% level of confidence.

– traceability of BIPM prototypes to the new international prototype of the kilogram be confirmed.

These conditions have been discussed and compared with the present capabilities of mass comparisons and with the propagation of uncertainties from the realization of the kilogram to mass standards of class $E_1$ [14]. In this work the recommendations are accepted reasonable, however if the realization uncertainties are larger than those required by the CCM, mass measurements with accuracies higher than $2.5 \times 10^7$ would no longer be available.

The modification of the SI yet to be carried out depends on the results obtained by many metrological institutes. The best results have been obtained by the NIST and the IAC, when the Planck constant and the Avogadro constant have been measured with the relative standard uncertainties $5 \times 10^{-8}$ [15] and $3 \times 10^{-8}$ [16], correspondingly.

Careful analysis of a set of fixed FPC should be performed so as to prevent any logical or measurement inconsistencies related with new definitions of SI units and take advantage of freedom in choosing of some definitions. For instance, it is desirable to use the fact concerning dependence of measurement accuracy on the measurement units involved. It is known, that the relative standard uncertainties of the masses of the electron, the proton, the neutron and the alpha particle measured in the kilograms are on the $10^{-7}$ level, while the same quantities measured in the unified atomic mass units (u) are on the $10^{-10}$ level.

It was proposed the following definitions of the basic SI units in terms of the FPC with the fixed values [2, 3].

The ground state hyperfine splitting frequency of the caesium-133 atom $\Delta\nu\,(^{133}Cs)_{hfs}$ is exactly 9 192 631 770 hertz, Hz.

The speed of light in vacuum $c$ is exactly 299 792 458 metre per second, m s$^{-1}$.

The Planck constant $h$ is exactly 6.626 069 3×10$^{-34}$ joule second, J s.

The elementary charge $e$ is exactly 1.602 176 487×10$^{-19}$ coulomb, C.

The Avogadro constant $N_A$ is exactly 6.022 141 79×10$^{23}$ reciprocal mol, mol$^{-1}$.

The Boltzmann constant $k$ is exactly 1.380 650 4×10$^{-23}$ joule per kelvin, J K$^{-1}$.

The luminous efficacy $K_{cd}$ of monochromatic radiation of frequency 540×10$^{12}$ Hz is exactly 683 lumen per watt, lm W$^{-1}$.

## 3. Some shortcomings of SI units redefinitions based on fixed FPC

The written above seven fixed constants can set the scale of the new SI, but if this set will be adopted entirely, some problems are capable of affecting adversely on progress of metrology and physics. For instance, the fixation of $h$ concurrently with $N_A$ leads to the fixation of a combination of the following constants: the speed of light in vacuum $c$, the molar mass constant $M_u$, the relative atomic mass of the electron, the fine-structure constant $\alpha$ and the Rydberg constant $R_\infty$.

Really, there is a relation between $h$ and $N_A$, what combines the mentioned above FPC measured with higher accuracy (~10$^9$) than the accuracies of $h$ and $N_A$ [15]:

$$\frac{R_\infty}{A_r(e)\alpha^2} = \frac{M_u c}{2hN_A}, \qquad (1)$$

$M_u$ is exactly 10$^{-3}$ kg/mol, $M_u = M(^{12}C)/12$, $M(^{12}C)$ is the molar mass of the carbon-12. The constants $c$ and $M_u$ is exactly known, while $A_r(e)$, $R_\infty$ and $\alpha$ are experimentally measured constants, so the concurrent $h$ and $N_A$ fixation leads to the additional non-physical constraint of their values.

If the kilogram is defined through the fixed $h$ value, then only two among $N_A$, $h$ and $M$ ($^{12}$C) may be defined independently, this point has been noticed also in Ref. [17]. It has been suggested to fix $N_A$ and $M(^{12}C)$, in this case the $h$ value remains free for a subsequent concordance in the family of electromagnetic quantities [12, 18].

It is known, that the high precision $h$ measurement is a complicated task, since there are many sources of systematical uncertainties influencing on a final result of such a complexly built device as a watt balance. So the determination of the fixed value of the Planck constant with the help of the only experimental device may cause to inaccurate consequences.

By way of illustration let us consider determination of the Planck constant and the Avogadro constant by the silicon spheres method and the watt balance method [19, 20, 21]. We know that the discrepancy on the $10^{-6}$ level for the relative difference of the values of these constants was obtained upon completion of the first stage of the Avogadro project; that is values of systematical uncertainties for each of these experiments were underestimated by itself. The relative difference of the results obtained in the NPL and in the NIST is about 0.3 ppm, therewith the NPL result is over the NIST result and is closer to the result obtained on the first stage of the Avogadro project [19, 21].

Besides, the use of watt balances and a feasible subsequent adoption of "the electronic kilogram", based on the electric quantum units of voltage and resistance, depends on a verification of the Josephson and von Klitzing relations, for instance, through devices for a single electron tunneling. Then the closure of the so-called "quantum triangle" will be possible, along with an adoption of consistent system of practical electrical units [22, 23, 24]. The point is that the Josephson and von Klitzing relations, which confirm experimentally on the levels ~ $10^{-7}$ for the Josephson relation and ~ $10^{-9}$ for the von Klitzing relation, are justified theoretically in the semi classical approximation for a sufficiently great number of electrons, however this approximation does not work in contemporary nanotechnological devices. It is expected that independent confirmation of these relations will be perform upon completion of producing of a quantum current standard based on direct counting of an electron amount which goes throughout a given section of a transmitter per second.

Independent measuring of $h$ with the help of a watt balance after a closure of the "quantum triangle" will be an additional check of the relation: $K_J^2 R_K = 4/h$. This relation has no proof in the strict sense; therefore it is possible that some corrections $\varepsilon_J$ and $\varepsilon_K$ exist:

$$K_J^2 R_K (1 - 2\varepsilon_J - \varepsilon_K) = 4/h, \qquad (2)$$

where $\varepsilon_J$ and $\varepsilon_K$ are corrections, which take into account possible deviations of $K_J$ and $R_K$ from the expressions $2e/h$ and $h/e^2$, respectively.

It has thrown doubt on the necessity of the fixation of the Planck constant for the redefinitions of SI units. If the new definition of the kilogram with the fixed $h$ value be adopted, then the basis of this definition will be the complicated electromechanical device with many sources of systematical uncertainties – the watt balance – rather than a nature invariant. Alternatively, if the new definition of the kilogram with the fixed $N_A$ value be adopted, the precise method of the Planck constant determination will be of use, which depends on the values of the electron mass, the Rydberg constant ant the fine-structure constant through the relation (1). In this case the watt balance method will become the method for the precise measurement of $K_J^2 R_K$, that will lead to the further progress in the theory of the macroscopic quantum Josephson and von Klitzing effects.

At the present time the CODATA $N_A$ value is determined with the help of the $h$ value measured with the NIST watt balance. The $h$ and $N_A$ values are equal to the following quantities with the relative standard uncertainties $5.0 \times 10^{-8}$ [15]:

$$h = 6.62606896(33) \times 10^{-34} \, J \, s, \qquad (3)$$

$$N_A = 6.02214179(30) \times 10^{23} \, mol^{-1}. \qquad (4)$$

The freedom of choosing of dimensions of the basic units based on fixed FPC values is bounded by the condition of the maximal consistency with the dimensions now in use. So, a great many FPC values should not be fixed for redefinitions of SI units. It is best to redefine SI units with a few fixed FPC at achievement of desired purposes, namely, the unity, the precision and the stability of measurements in the framework of the new SI. To do this requires criteria for choosing of FPC with fixed values. In the next section such criteria will be suggested along with a possible set of a few fixed FPC for redefinition of SI units.

**4. Criteria for choosing of FPC with fixed values**

Let us consider what requirements must be met physical quantities with fixed values for redefinition of SI units. It might appear at first sight that such physical quantities do not exist, since any physical quantity is measured with some uncertainties. However, this is not always the case; because of there is an arbitrariness of definitions of measurement units, which is constrained only by the succession condition for consistency with historically completed dimensions of units.

Values of only certain number of FPC or combinations of FPC can be fixed if their constancy is confirmed with high accuracy under any exposure. An example of this type of fixation is the fixation of the speed of light in vacuum. The experimental verification of the special theory of relativity is the physical justification of this fixation, while the metrological basis consists in higher accuracy of measurement of time separations as compared with lengths. It is not inconceivable that space or time variations of the speed of light will be discovered with better measurement arrangements. However, at present it is valid to say that the speed of light is the absolute metrological invariant at the modern theoretical and experimental level. The value of the speed of light can be fixed due to the specific freedom at a definition of the length unit, namely, the metre ("the light metre"). Non-contradictory definitions of other units along these lines are bounded by the evident condition of agreement on a workable level of accuracy between values of old and new units' dimensions.

By this means a few independent quantities or their combinations can be fixed in the SI framework, and the invariance of these quantities must have by far a sound theoretical or experimental support.

What quantities do fit these conditions? The charges and masses of existing molecules, atoms and elementary particles fit these conditions. A great body of data shows identity of these physical systems in the present era and in the past. Thus, the absolute value of the electron charge and the mass value of the carbon atom $^{12}$C can be fixed. The fixation of the mass value of the carbon atom provides in fact the basis for the etalon of the microscopic mass unit.

Main criteria for choosing of fixed FPC for a redefinition of SI units may be formulated as follows. One should take into account two things when fixed a physical quantity for a redefinition of a SI unit. First, is any nature or conceivable invariant with the same physical dimensions as the considered SI unit available? Second, is any FPC with the physical dimensions which differ from the physical dimensions of the SI unit only by any power of the time/frequency available? The last criterion appears explicable due to the unprecedented measurement precision (~$10^{16}$ and better) of the time/frequency quantities as compared to other quantities.

Under these criteria, the dimensionless value of the Avogadro constant $\{N_A\}$ can be fixed for the redefinition of the amount of substance unit. Moreover, it is possible to pick out the fixed number of the carbon atoms with the total mass which is equal to the IPK mass within the definite uncertainties. By this means the etalon of the macroscopic mass unit based on the carbon atom mass is to be defined.

It should be noted that in many papers the definitions of the mass and amount of substance units have been chosen independently of one another. In the present paper we use the principle of the minimum of the fixed FPC, so the definitions of the mass and amount of substance depend of one another through the molar mass of the carbon-12 atom. A procedure of a realization of new definitions of SI units based on the minimal set of the fixed FPC is considered in the next section.

From the above reasoning it is clear that needless to fix the Planck constant. This constant is best included in the set of constants with values adjusted to results of different experiments. Besides, the fixation of the Planck constant concurrent with the Avogadro constant leads to the additional non-physical constraint of the $A_r(e)$, $R_\infty$ and $\alpha$ values, as it indicated in Sec. 3. This constraint can bring in contradiction with future data of precise measurements of these constants. In addition, the watt balance method can become the method of determination of the corrections $\varepsilon_J$ and $\varepsilon_K$ to the Josephson and von Klitzing relations.

## 5. Peculiarities of realization of new definitions of SI units based on fixed FPC

Suggestions concerning redefinitions of four SI units with the help of fixed FPC are based on the same concept which was used for the definition of the light metre in 1983. At present, it is suggested to fix four FPC, that is the $h$, $e$, $k_B$ and $N_A$ values will be defined with zero uncertainty, in order to determine properly the kilogram, the ampere, the kelvin and the mol.

However, in 1983 when the light metre was defined, the measurements of the frequency and the speed of light had been performed, which had the precision more than order of magnitude higher than the length measurements. This situation led to the opportunity to redefine the speed of light. Things get worse for the Planck constant $h$, the elementary charge $e$, the Boltzmann constant $k_B$ and the Avogadro constant $N_A$ with the own situation for each constant, this raises the question about the premature fixation all constants in one stroke.

Actually, it is desirable to determine the minimum set of the FPC in order to provide the unity and the high precision of measurements. It is ideal in so doing, if new definitions of measurement units would be operational, that is to say, would directly reproducible in actual practice. For instance, uncertainties of high precision length measurements are determined in practice with uncertainties of positions of interference strips of laser beams, whereas when evaluating uncertainties of length measurements are estimated with uncertainties of frequency measurements. An application of new definitions of some SI units may be in perfect analogy with the application of the light metre, so it is needed to work out in detail the procedure of realization of new definitions of SI units.

As indicated above, the minimum set of fixed FPC is best suited for redefinitions of SI units. The suggestion of fixation as many constants as there are SI units depends, possibly, upon the fact that some authors intend to solve at the same time a few problems, namely, to introduce a new definition of the mass unit and to legalize in the SI framework using of the practical electric units based on the quantum Hall effect and the Josephson effect. However, at present there is not the complete theory of these effects, which allows, for example, evaluating corrections depending on the test material and extreme environmental influences. It is very desirable to test with high accuracy the relations, which contain the Planck, Josephson, von Klitzing constants and the elementary charge.

A remark is in order. It is known that at present the introduction of the fixed Boltzmann constant is not warranted in practice. Existing accuracy of temperature measurements is reasonable for practical measurements and will not considerably increase in the coming years. Furthermore, the precise Boltzmann constant value must be determined from a few experiments and be consistent with values of other constant of molecular physics [25, 26].

On this basis it is reasonable safe to suggest that it is justified only the redefinition of the ampere based on the fixed elementary charge and the redefinitions of the kilogram and the mol based on the fixed Avogadro constant and the molar mass of the carbon-12 atom (quantities at the fixed Avogadro constant $N_A^*$ are denoted with an asterisk further on).

Let us call attention to the following peculiarity of choosing the fixed $N_A^*$ value. In the SI now in operation it historically happened that the mass unit and the amount of substance unit are connected to one another. In the present paper this connection is suggested to retain in the new

SI. Thus $\{N_A{}^*\}$ and the fixed number of the carbon atoms $\{N_{kg}{}^*\}$, which defines the macroscopic mass unit in the new SI (kilogram*), must be connected by the following relation: $\{N_A{}^*\} = 0.012\{N_{kg}{}^*\}$.

Attention must be centred on the following definition of the mass unit: *the kilogram\* is the mass of the $\{N_A{}^*\}/0.012$ free carbon $^{12}C$ atoms in the rest and in the ground quantum state.*

The considerable advantage of the definition of the mass unit under discussion lies in the fact that this definition is correlated well with the definition of the amount of substance unit, namely, *the mol\* consists of the $\{N_A{}^*\}$ structure units of the given substance.*

As evident from these definitions, the value $\{N_A{}^*\}$ must be divisible by 0.012. In this case the kilogram* and the gram* will contain the integer number of the carbon $^{12}C$ atoms, such definitions are convenient to use. Hence the nature invariant, the mass of the carbon $^{12}C$ atom, is the basis of the new definition of the kilogram. The main problem is of choosing of the specific fixed value $\{N_A{}^*\}$ on the acceptable interval in such a way as to make the relative difference of the existing IPK value $\{K\}$ and the new kilogram value $\{K^*\}$ of the order of $10^{-8}$. Then the new molar mass of the carbon-12 atoms $M(^{12}C)^*$ and the new molar mass constant $M_u{}^*$ are expressible through the kilogram* and the mol* in much the same way as in the old SI.

$$M(^{12}C)^* = 12\times10^{-3} \text{ kg}^*/\text{mol}^*, \quad M_u{}^* = 10^{-3} \text{ kg}^*/\text{mol}^*. \qquad (5)$$

This version of new definitions of the kilogram, the mol, the molar mass of the carbon-12 atoms and the molar mass constant is very convenient, since it does not disturb the existing situation with evaluations of results of mass and molar mass measurements in chemistry [27, 17, 18].

As an example let us consider a specific fixation of the Avogadro number $\{N_A{}^*\}$. At the moment the $N_A$ values, which have been obtained by the IAC [16], should be used.

$$\{N_A\} = (6.02214066 \div 6.02214102) \times 10^{23} \qquad (6)$$

Now, take into account the relation $\{N_A^*\} = 0.012\{N_{kg}^*\}$, we pick out number, which is divisible by 0.012 and is nearest to the middle of the interval. This number is $6.02214084 \times 10^{23}$, so the fixed Avogadro number is

$$\{N_A^*\} = 6.02214084 \times 10^{23}.  \qquad (7)$$

The presented above definitions of the mass unit and the amount of substance unit based on the fixed value of the Avogadro constant $\{N_A^*\}$ will maintain succession with the definitions now in operations of the kilogram and the mol, as well as will not disturb the existing metrological chains of passing of dimensions of the mass and amount of substance units and the current practice of mass and molar mass measurements [14, 15, 17, 18].

The prevailing view today is that the new definition of the kilogram should be based on the fixed Planck constant $h$, which is the fundamental constant of quantum physics, much as $c$ is the fundamental constant of the special theory of relativity. Besides that, the redefinition of the ampere by the fixation of the elementary charge $e$, while the quantities $2e/h$ and $h/e^2$ have the exact values, will provide advantages for the metrology of electric measurements and will carry the practical electric units $\Omega_{90}$ and $V_{90}$ in the class of SI units. However, as mentioned above, the fixation of the Planck constant h, together with the relations $K_J = 2e/h$ and $R_K = h/e^2$, may be a non-correct procedure. But, the fixed values of the Avogadro constant, the $^{12}C$ atom mass and the elementary charge at distant interactions are the absolute metrological invariants at any time and under any conditions.

There are strong grounds for believing that the new definitions of the mass unit and the amount of substance unit will be based on the fixed Avogadro constant and the $^{12}C$ atom mass [12, 18]. This logical consistent method leads to the descriptive and sensible definitions of the kilogram and the mol, which are compatible with the current definitions of these units and the practice of mass and amount of substance measurements.

We finally arrive at the following, essentially uniquely determined, set of five definitions of the SI units based on the fixed FPC.

The second is the unit of time, which is equal to the 9192631770 periods of radiation originated at the ground state hyperfine splitting transition of the caesium-133 atom at rest at a temperature of 0 K.

The metre is the unit of length, which is equal to the distance travelled by light in vacuum in one second divided by 299792458.

The kilogram is the unit of mass, which is equal to the mass of $5.0184507 \times 10^{25}$ free carbon-12 atoms at rest in the ground state.

The mol is the unit of amount of substance, which contains $6.02214084 \times 10^{23}$ specified elementary entities.

The ampere is the unit of electric current, such that the value of the elementary charge is equal to $1.602176487 \times 10^{-19}$ ampere second.

It is important that the kelvin, the unit of thermodynamic temperature, is the physical quantity, which is intimately associated with the properties of the macroscopic bodies and cannot be defined with the help of quantum and numerical invariants. So in the framework of the considered approach the kelvin is not desirable to define through the fixed Boltzmann constant. Much remains to be done for the best possible redefinition of the temperature unit.

## 6. Summary

Considerable recent attention has been focused on the possibility of the redefinition of four base units: the kilogram, the ampere, the kelvin and the mol, with the help of the fixed values of the Planck constant, the Avogadro constant, the Boltzmann constant and the elementary charge. The main problem is choosing some optimum set of FPC, whose values should be fixed for the redefinitions of the SI units, besides the problem of increasing of precision for constants' measurements.

According to the adopted recommendations and resolutions of the CIPM and the 23[rd] CGPM, it is proposed to the National Metrological Institutes to continue for increasing efforts in realizations of experimental and theoretical works aimed to new definitions of SI units based on fixed FPC. Take into account different opinions on this subject and difficulties of increasing the accuracy of constants' values, the SI reform will most likely take time up to 2019 inclusive.

In the present paper the criteria on choosing of the optimum set with the fixed FPC for the redefinition of the SI units are suggested. Among which are the criterion of existing of nature invariants with the same dimensions as with the dimensions of defined units and the criterion of existing of FPC with dimensions differed from the dimensions of defined units by any power of the time.

In the paper we considered also the minimal optimum FPC set that is consistent with the criteria. The set comprises the speed of light, the constant of the ground state hyperfine transition of the caesium-133 atom, the Avogadro constant, the mass of the carbon-12 atom and the absolute magnitude of the electron charge. The redefinition of the kelvin in the framework of this approach is impossible. Additional efforts are needed to improve the definition of this unit. It is suggested also do not fix the values of the Planck, Josephson and von Klitzing constants and to continue investigations to produce a quantum current standard and to verify the Josephson and von Klitzing relations on the high precision level.